\documentclass[12pt,showkeys,showpacs] {revtex4}
\usepackage{graphicx}
\begin{document}

\title[Short Title]{Improving the security of multiparty quantum secret splitting and quantum state sharing}
\author{Hong-Fu Wang}
\author{Xin Ji}
\author{Shou Zhang\footnote{E-mail: szhang@ybu.edu.cn}}
\affiliation{Department of Physics, College of Science, Yanbian
University, Yanji, Jilin 133002, PR China}

\begin{abstract} A protocol for multiparty quantum secret
splitting (MQSS) with an ordered $N$ Einstein-Podolsky-Rosen (EPR)
pairs and Bell state measurements is recently proposed by Deng {\rm
et al.} [Phys. Lett. A 354(2006)190]. We analyzed the security of
the protocol and found that this protocol is secure for any other
eavesdropper except for the agent Bob who adopts
intercept-and-resend attack. Bob can obtain all the information of
Alice's alone without being found. We also propose an improved
version of the MQSS protocol.


\end{abstract}

\maketitle Quantum mechanics principles supplied many interesting
applications in the field of quantum information in the last
decade. The interesting aspect of employing quantum mechanics for
quantum secret sharing (QSS) is that it allows the unconditionally
secure distribution of the information between the participants.
It is a generalization of classical secret sharing \cite
{BACWNY9670, JFCTCPL9704} to a quantum scenario \cite
{MVAPRA9959}. Since the pioneering work presented by Hillery {\rm
et al.} in 1999 by using three-particle and four-particle
Greenberger-Horne-Zeilinger (GHZ) states \cite {MVAPRA9959}, a lot
of QSS protocols \cite{AMNPRA9959, AJHPRA01, GGPLA03, AMLPRL04,
YXSPLA05, FHGPLA05, FGHPLA05, LCPRA05, AMLPRA05, WHNPRA01,
SBPRA00, HFCPRA03, LGFJPRA04, ZJZPLA05, FTPRA05, FXCPHPLA06} have
been proposed in both theoretical and experimental aspects. QSS is
an useful tool in the cryptographic applications. Suppose that
Alice wants to send a message to her agents, Bob and Charlie, who
are at remote places to deal with her business. However Alice
doubts that one of them may be dishonest and she does not know who
the dishonest one is, but she knows that if the two of them
coexist, the honest one can keep the dishonest one from doing any
damage. To prevent the dishonest man from destroying the business,
Alice splits the secret messages into two encrypted parts and
sends each one a part so that neither individual is able to obtain
all of the original information unless they collaborate.

Recently, Deng  {\rm et al.} \cite{FXCPHPLA06} proposed a
multiparty quantum secret splitting and quantum state sharing
protocol (hereafter called Deng's protocol) for splitting a secret
message among three parties, say Alice, Bob, and Charlie with EPR
pairs following some ideas in Refs. \cite{ZJZPLA05, FGXPRA0368}.
In this paper, we show that Deng's protocol can be eavesdropped by
a dishonest agent Bob who adopts intercept-and-resend attack, and
Bob can obtain the whole secret messages of Alice's alone without
being found. Moreover, we propose an improved version of Deng's
protocol.

Let us briefly review Deng's protocol. For convenience, we only
discuss the simple case in which there are three parties, the Boss
Alice, two agents, Bob and Charlie. The principle for the case of
multiparty is the same as this simple one with just a little
modification or not. In Deng's protocol, the agent Bob prepares an
ordered $N$ EPR polarization photon pairs in the same quantum
state:
$|\psi^-\rangle_{AC}=\frac{1}{\sqrt{2}}({|0\rangle_A}{|1\rangle_C}-{|1\rangle_A}{|0\rangle_C})$.
The $N$ ordered EPR pairs are denoted with \{[${\rm P}_1$($A$),
${\rm P}_1$($C$)], [${\rm P}_2$($A$), ${\rm P}_2$($C$)], \ldots,
[${\rm P}_N$($A$), ${\rm P}_N$($C$)]\}. Bob takes one photon from
each EPR pair to form an ordered EPR partner photon sequence, say
[${\rm P}_1$($A$), ${\rm P}_2$($A$), \ldots, ${\rm P}_N$($A$)],
called the $S_A$ sequence. The remaining EPR partner photons
compose another EPR partner photon sequence [${\rm P}_1$($C$),
${\rm P}_2$($C$), \ldots, ${\rm P}_N$($C$)], which is called the
$S_C$ sequence. Then Bob first sends the $S_A$ sequence to Alice
and keeps the $S_C$ sequence. Alice picks out a sufficiently large
subset of photons from the sequence $S_A$ for the eavesdropping
check of the transmission. The check can be completed with the
following procedures: (a) Alice tells Bob which photons he has
chosen and Bob picks out the correlated photons in the sequence
$S_C$. (b) Bob randomly chooses the measuring basis (MB) $Z$ or
$X$ to measure the chosen photons. (c) Bob tells Alice which MB he
has chosen for each photon and his measurement results. (d) Alice
uses the same MBs as Bob to measure the corresponding photons and
checks the eavesdropping with the results of Bob's. If the error
$\varepsilon_1$ is small, Alice and Bob can conclude that there
are no eavesdropping in the line, this is the first eavesdropping
check. After that, Bob randomly chooses one of the four local
unitary operations $U_i$ (${U_0}\equiv I$, ${U_1}\equiv\sigma_x$,
${U_2}\equiv i{\sigma_y}$, ${U_3}\equiv\sigma_z$, $I$ is the
identity operator, $\sigma_i$ are the Pauli operators, and $i=0,
1, 2, 3$) to encrypt each of the photons in the $S_C$ sequence,
say $U_B$, and then sends the $S_C$ sequence to Charlie. Alice and
Charlie analyze the error rate $\varepsilon_2$ of the transmission
of $S_C$ sequence, similar to that for $S_A$ sequence. The
difference is that Alice should require Bob to publish his unitary
operations on the sample photons before she accomplishes the
eavesdropping check. This is the second eavesdropping check. If
the error $\varepsilon_2$ is small, Alice and Charlie continue to
the next step; otherwise they have to abandon their transmission.
Alice selects a subset of photons as the samples for eavesdropping
check and chooses one of the four unitary operations randomly on
each sample photon. For other photons in the $S_A$ sequence
(except for those for eavesdropping check), Alice encodes her
secret message $M_A$ on them with the four unitary operations, and
then sends the sequence $S_A$ to Charlie. Charlie performs the
Bell measurements on the EPR photon pairs and reads out the
combination of operations done by Alice and Bob. Alice and Charlie
finish the checking of error rate of the samples selected by
Alice, if the error is small, Alice tells Bob and Charlie to
collaborate for reading out the message $M_A$.

As pointed out in the Refs. \cite{MVAPRA9959, AMNPRA9959}, if a
dishonest agent in an MQSS cannot eavesdrop the secret messages
without disturbing the quantum system, any eavesdropper can be
found out. In this way, the main goal for the security of an MQSS
protocol is simplified to keep the dishonest agent from
eavesdropping the secret messages. In Deng's protocol, any
eavesdropper cannot get the secret messages of Alice's except for
the dishonest agent Bob who adopts intercept-and-resend attack.
Bob can eavesdrop the secret messages freely without being found.
Now we discuss the intercept-and-resend attack adopted by Bob in
detail.

In order to check for the eavesdropping there are two main steps:
Alice-Bob check and Alice-Charlie check, i.e., the step(4) and
step (6) in Deng's protocol. We describe the intercept-and-resend
attack from step (5) after Alice and Bob have finished the first
eavesdropping check and no eavesdropping is found in Alice-Bob
line. Instead of sending to Charlie the $S_C$ sequence, Bob could
instead prepare a new ordered $M$ ($M<N$) EPR pairs in the same
quantum state: $|\psi^-\rangle_{A^\prime
C^\prime}=\frac{1}{\sqrt{2}}({|0\rangle_{A^\prime}}{|1\rangle_{C^\prime}}-{|1\rangle_{A^\prime}}{|0\rangle_{C^\prime}})$
according to the numbers of the remaining EPR pairs in $S_A$
sequence after the first eavesdropping check. Bob takes one photon
from each new EPR pair to form an new ordered EPR partner photon
sequence, say [${\rm P}_1$($A^\prime$), ${\rm P}_2$($A^\prime$),
\ldots, ${\rm P}_M$($A^\prime$)], called the $S_{A^\prime}$
sequence. The remaining EPR partner photons compose another EPR
partner photon sequence [${\rm P}_1$($C^\prime$), ${\rm
P}_2$($C^\prime$), \ldots, ${\rm P}_M$($C^\prime$)], which is
called the $S_{C^\prime}$ sequence. Bob randomly chooses one of
the four local unitary operations $U_i$ to encrypt each of the
photons in the sequence $S_{C^\prime}$, say $U_{B^\prime}$, and
then he sends the sequence $S_{C^\prime}$ to Charlie and keeps the
$S_{A^\prime}$ sequence. Moreover, note that Bob also keeps the
original sequence $S_C$.

After Charlie confirms Alice he has received the $S_C$ sequence
(in fact, the sequence he receives is the $S_{C^\prime}$ sequence)
from Bob, Alice announces which photons will be used to check for
eavesdropping, Bob performs entanglement swapping between those
photons in sequence $S_C$ and the corresponding photons in the
sequence $S_{A^\prime}$. As a result, Alice and Charlie perfectly
share some new EPR pairs consisting of the photons of sequence
$S_A$ and $S_{C^\prime}$ for those positions that are going to be
tested for eavesdropping. Now Bob announces those unitary
operations corresponding to his measurement results of the
entanglement swapping. Alice and Charlie will find that all the
photons that they test reveal no eavesdropping. In this way, Alice
and Charlie confirm that there are no eavesdropping in the line.
Alice selects a subset of photons as the samples for eavesdropping
check and chooses one of the four unitary operations randomly on
each sample. For other photons in the $S_A$ sequence (except for
those for eavesdropping check), Alice encodes her secret message
$M_A$ on them with the four unitary operations, and then sends the
sequence $S_A$ to Charlie.

Bob intercepts the sequence $S_A$ when Alice sends the sequence
$S_A$ to Charlie, and performs the Bell measurements on the EPR
photon pairs, i.e., on the photons of sequences $S_A$ and $S_C$,
and read out the unitary operations that Alice has performed on
the photons of sequence $S_A$, i.e., Alice's secret messages. Then
Bob performs the correct unitary operations as Alice on the
photons of the $S_{A^\prime}$ sequence and sends the
$S_{A^\prime}$ sequence (except for those photons for
eavesdropping check) to Charlie. As a result, Charlie possesses
another ordered $M^\prime$ ($M^\prime<M$) EPR pairs.

Charlie confirms Alice he has received the $S_A$ sequence (in
fact, the sequence he receives is the $S_{A^\prime}$ sequence),
and performs the Bell measurements on the photons $A^\prime$ and
$C^\prime$. Alice announces which photons will be used to check
for eavesdropping, she first requires Bob to publish the unitary
operations done on the correlated photons in the sequence $S_C$,
and then she requires Charlie to tell her the results of the Bell
measurements. Bob announces the unitary operations that he has
performed on the photons of the sequence $S_{C^\prime}$. Alice
analyzes the error rate of the sample photons with Charlie's Bell
measurement results, Bob's operations, and her own operations. As
a result, Alice will find that there are no eavesdropping in the
line, and tell Bob and Charlie to collaborate for reading out the
message $M_A$. In this way, Bob can completely obtain Alice's
secret messages alone without being found. Finally, if Bob wants
to let Charlie obtain the secret messages of Alice's, he announces
the accurate unitary operations $U_{B^\prime}$ he has performed on
the photons in the sequence $S_{C^\prime}$. Otherwise, Charlie
will not obtain the accurate secret messages of Alice's.

So far we have pointed out that Deng's protocol is not secure when
the agent Bob is dishonest in the case of three-party QSS. Bob can
eavesdrop the secret messages of Alice's alone without being found
during the process of eavesdropping check by adopting
intercept-and-resend attack. The similar case happens for MQSS.
For improving the security of Deng's protocol \cite{FXCPHPLA06},
the three parties must have the ability to keep the eavesdropper
from eavesdropping the secret messages. The Deng's MQSS protocol
is secure if Alice and Charlie can prevent the agent Bob who
adopts intercept-and-resend attack from eavesdropping the secret
messages. Now let us describe the modified Deng's MQSS protocol in
the case of $M$ agents as follows.

Step 1: Alice prepares an ordered $N$ EPR polarization photon
pairs in the same quantum state:
$|\psi^-\rangle_{AT}=\frac{1}{\sqrt{2}}({|0\rangle_A}{|1\rangle_T}-{|1\rangle_A}{|0\rangle_T})$.
She divides the photons into two sequences: [${\rm P}_1$($A$),
${\rm P}_2$($A$), \ldots, ${\rm P}_N$($A$)], [${\rm P}_1$($T$),
${\rm P}_2$($T$), \ldots, ${\rm P}_N$($T$)], which are called
$S_A$ sequence and $S_T$ sequence, respectively. Then Alice sends
the $S_T$ sequence to Bob.

Step 2: After receiving Bob's confirmation of receiving the $S_T$
sequence, Alice picks out a sufficiently large subset of photons
from the sequence $S_A$ for the eavesdropping check of the
transmission.

The check can be completed with the following procedures: (a1)
Alice tells Bob which photons he has chosen and Bob picks out the
correlated photons in the sequence $S_T$ (a2) Bob randomly chooses
the measuring basis (MB) $Z$ or $X$ to measure the sample photons.
(a3) Bob tells Alice the MB he has chosen for each photon and the
results of his measurements. (a4) Alice uses the same MBs as Bob
to measure the corresponding photons in $S_A$ sequence and checks
the eavesdropping with the results of Bob's. If the error rate is
small, Alice and Bob can conclude that there are no eavesdropping
in the line. Alice and Bob continue to perform step 3; otherwise
they have to discard their transmission and abort the
communication.

Step 3: Bob randomly chooses some photons from $S_T$ sequence as
sample photons and performs a Hadamard transformation ($H $
transformation) on each of the sample photons. The $H$
transformation can transform each qubit as
\begin{eqnarray}\label{e1}
&{H}{|0\rangle}=\frac{1}{\sqrt{2}}(|0\rangle+|1\rangle),&\cr\cr&{H}{|1\rangle}=\frac{1}{\sqrt{2}}(|0\rangle-|1\rangle).
\end{eqnarray}
For the other photons in $S_T$ sequence, Bob randomly chooses one of the four local unitary operations $U_i$
$(i=0, 1, 2,3)$ to encrypt each of the photons, say $U_B$, and
then he sends the sequence $S_T$ to Charlie.

Step 4: Charlie confirms Alice and Bob that he has received the
$S_T$ sequence, then Bob announces the position of the sample
photons to Alice and Charlie. Alice and Charlie analyze the error
rate of the transmission of sequence $S_T$, similar to that in
step 2. The difference is that Charlie should first perform a $H $
transformation on each of the sample photons before she
accomplishes the eavesdropping check. If the transmission is
secure, Charlie randomly chooses some photons from $S_T$ sequence
as sample photons and performs a $H$ transformation on each of the
sample photons. For the other photons in $S_T$ sequence, Charlie
randomly performs one of the four unitary operations on each
photon, say $U_C$, and then sends the sequence $S_T$ to the next
agents.

Step 5: After repeating the step 4 $M-3$ times, the $S_T$ sequence
is received securely by the ($M-1$)th agent, say Yang. The
($M-2$)th agent announces the position of the sample photons to
Alice and Yang, then Alice and Yang analyze the error rate of the
transmission of sequence $S_T$, similar to that in step 4. The
difference is that Alice should require the foregoing $M-2$ agents
except for the ($M-2$)th agent to publish their unitary operations
on the sample photons before she accomplishes the eavesdropping
check. If the transmission is secure, Yang randomly chooses some
photons from $S_T$ sequence as sample photons and performs a $H$
transformation on each of the sample photons. For the other
photons in $S_T$ sequence, Yang randomly performs one of the four
unitary operations on each photon, say $U_Y$, and then sends the
sequence $S_T$ back to Alice.

Step 6: Alice confirms Yang that she has received the sequence
$S_T$, then Yang announces the position of the sample photons to
Alice. Alice picks out the sample photons from $S_T$ sequence and
the correlated photons from $S_A$ sequence for the eavesdropping
check of transmission. This can be completed as follows: (b1) The
foregoing $M-1$ agents except for Yang tell Alice the operations
that they have performed on the sample photons. (b2) Alice
performs a Hadamard transformation on each of the sample photons
and then performs the Bell measurements on the EPR photon pairs
after she knows the $M-2$ agents' operations. (b3) If Alice's Bell
measurement results coincide with the $M-2$ agents' operations,
Alice continues to the next step; otherwise Alice has to discard
her transmission and abort the communication.

Step 7: Alice encodes her secret messages $M_A$ on the photons in
$S_A$ sequence with the four unitary operations, say $U_A$. Then
she prepares a sufficiently large number of single photons, which
we call as checking photons, in one of the four quantum states
that constitute two bases $\{|0\rangle, |1\rangle\}$ and
$\{|+\rangle, |-\rangle\}$, where
$|+\rangle=\frac{1}{\sqrt{2}}(|0\rangle+|1\rangle)$ and
$|-\rangle=\frac{1}{\sqrt{2}}(|0\rangle-|1\rangle)$. Alice mixes
these checking photons with the $S_A$ sequence and $S_T$ sequence
respectively, and firstly sends the $S_T$ sequence to the last
agent, Zach.

Step 8: After receiving Zach's confirmation of receiving the $S_T$
sequence, Alice tells Zach the position and the preparation basis
of the checking photons. Zach picks out the corresponding checking
photons and measures them in those basis and compares the results
with Alice publicly. If the error is small, Alice and Zach
continue to the next step; otherwise they have to abandon their
transmission.

Step 9: Alice sends the $S_A$ sequence to Zach. Alice and Zach
analyze the security of transmission of $S_A$ sequence, similar to
that in step 8. If the error is small, Zach continues to the next
step; otherwise they have to abandon their communication.

Step 10: Zach performs the Bell measurements on the EPR photon
pairs and reads out the combination of the operations done by
Alice, Bob, \ldots, and Yang, i.e., $U_Z=U_A\otimes
U_B\otimes\ldots\otimes U_Y$.

Step 11: If all the agents agree to collaborate, they can read out
the secret messages $M_A$, otherwise they will abandon the results
of the transmission.

Up to now, we have proposed an improved version of Deng's MQSS
protocol. The modified MQSS protocol is secure by randomly
performing the Hadamard transformation and introducing auxiliary
checking photons. In this way, if the eavesdroppers including the
agents attempt to eavesdrop the secret messages, their
eavesdropping behaviors will be detected when Alice compared the
results with the agents.

In summary, we analyzed the security of the MQSS protocol proposed
by Deng {\rm et al.} \cite{FXCPHPLA06} and found that this
protocol is secure for any other eavesdropper except for the agent
Bob who adopts intercept-and-resend attack. Bob can get all the
secret messages of Alice's without being detected. Finally, we
present a possible improvement of the MQSS protocol security via
randomly performing the Hadamard transformation and introducing
auxiliary checking photons. With these modifications, Deng's MQSS
protocol is secure not only against the outsider's eavesdropping
but also against the agent's eavesdropping.

\end{document}